\documentclass[%
 reprint,
 superscriptaddress,
 amsmath,amssymb,
 aps,
]{revtex4-2}

\usepackage{xcolor}
\usepackage{subcaption}
\usepackage{graphicx}
\usepackage{dcolumn}
\usepackage{bm}
\usepackage{amsmath}

\begin{document}

\preprint{APS/123-QED}

\title{Simulations of magnetic Bragg scattering in transmission electron microscopy}

\author{Justyn Snarski-Adamski}
 \email{justyn.snarski-adamski@ifmpan.poznan.pl}
\affiliation{%
Institute of Molecular Physics, Polish Academy of Sciences, M. Smoluchowskiego 17, 60-179 Poznań, Poland
}%
\author{Alexander Edstr\"{o}m}
\affiliation{Department of Applied Physics, School of Engineering Sciences, KTH Royal Institute of Technology, AlbaNova University Center, 10691 Stockholm, Sweden}

\author{Paul Zeiger}
\affiliation{
 Division of Materials Theory, Department of Physics and Astronomy, Uppsala University, Box 516, SE-751 20 Uppsala, Sweden
}%
\author{Jos\'{e} \'{A}ngel Castellanos-Reyes}
\affiliation{
 Division of Materials Theory, Department of Physics and Astronomy, Uppsala University, Box 516, SE-751 20 Uppsala, Sweden
}%
\author{Keenan Lyon}
\affiliation{
 Division of Materials Theory, Department of Physics and Astronomy, Uppsala University, Box 516, SE-751 20 Uppsala, Sweden
}%
 \author{Miros\l{}aw Werwi\'{n}ski}
\affiliation{%
Institute of Molecular Physics, Polish Academy of Sciences, M. Smoluchowskiego 17, 60-179 Poznań, Poland
}%
\author{J\'{a}n Rusz}
  \email{jan.rusz@physics.uu.se}
\affiliation{
 Division of Materials Theory, Department of Physics and Astronomy, Uppsala University, Box 516, SE-751 20 Uppsala, Sweden
}%

\begin{abstract}
We have simulated the magnetic Bragg scattering in transmission electron microscopy in two antiferromagnetic compounds, NiO and LaMnAsO. This weak magnetic phenomenon was experimentally observed in NiO by Loudon \cite{loudon_antiferromagnetism_2012}. We have computationally reproduced Loudon's experimental data, and for comparison we have performed calculations for the LaMnAsO compound as a more challenging case, containing lower concentration of magnetic elements and strongly scattering heavier non-magnetic elements. We have also described thickness and voltage dependence of the intensity of the antiferromagnetic Bragg spot for both compounds. We have considered lattice vibrations within two computational approaches, one assuming a static lattice with Debye-Waller smeared potentials, and another explicitly considering the atomic vibrations within the quantum excitations of phonons model (thermal diffuse scattering). The structural analysis shows that the antiferromagnetic Bragg spot appears in between (111) and (000) reflections for NiO, while for LaMnAsO the antiferromagnetic Bragg spot appears at the position of the (010) reflection in the diffraction pattern, which corresponds to a forbidden reflection of the crystal structure. Calculations predict that the intensity of the magnetic Bragg spot in NiO is significantly stronger than thermal diffuse scattering at room temperature. For LaMnAsO, the magnetic Bragg spot is weaker than the room-temperature thermal diffuse scattering, but its detection can be facilitated at reduced temperatures.
\end{abstract}

\maketitle


\section{\label{sec:Introduction}INTRODUCTION}


The rapid development in nanoengineering of magnetic materials calls for characterization methods at high spatial resolutions describing magnetic phenomena down to the atomic scale. Transmission electron microscopy (TEM) is among the natural choices for such high-resolution characterization of materials.
There are many variations of TEM techniques, including high resolution TEM (HRTEM)~\cite{rose_optics_2008}, differential phase contrast (DPC)~\citep{shibata_differential_2012}, elemental mapping using electron energy-loss spectroscopy (EELS) \cite{pennycook_atomic-resolution_2009}, scanning transmission electron microscopy (STEM)~\citep{leapman_scanning_1986}, and Lorentz STEM \citep{grundy_lorentz_1968, harada_lorentz_2022}. The study of the magnetic properties of materials can be done using various other techniques, including electron magnetic circular dichroism (EMCD) \citep{schattschneider_detection_2006}, which is a special case of EELS, magnetic electron holography \citep{thomas_electron_2008}, and recently also magnetic DPC at atomic scale \cite{edstrom_quantum_2019,krizek_atomically_2022,kohno_real-space_2022}, among others.

The electron beam consists of moving charged particles, i.e. an electrical current, which is susceptible to the electric and magnetic fields in the sample. By utilizing this interaction, it can be used for imaging the magnetic structure of materials. In absence of dynamical effects, theoretical understanding of imaging the magnetic structure in DPC method is based on Ehrenfest’s theorem \citep{muller_atomic_2014, lubk_differential_2015}. 


With the miniaturization of magnetic technology, there has been recent interest in thin film materials, especially antiferromagnetic materials (AFM) with collinear magnetic moments in light of their potential spintronic applications \cite{jungwirth_antiferromagnetic_2016,wadley_electrical_2016}. 
In Loudon's work \cite{loudon_antiferromagnetism_2012}, it was shown that observation of an antiferromagnetic Bragg spot in TEM is possible in NiO thin films. 
The antiferromagnetic reflection of magnetic Bragg scattering was found to be 10$^4$ times less intense than Bragg peaks derived from the structure and was found to clearly stand out from the TDS background. Using a two-beam model, Loudon furthermore estimated the oscillation period of the magnetic reflection to be about 236~nm. These results call for an investigation using a more elaborate computational model in order to understand the visibility of antiferromagnetic Bragg spots as a function of thickness and non-zero temperature.

\begin{figure}
   \centering
     \begin{subfigure}[b]{0.44\textwidth}
        \centering
        \includegraphics[width=\textwidth]{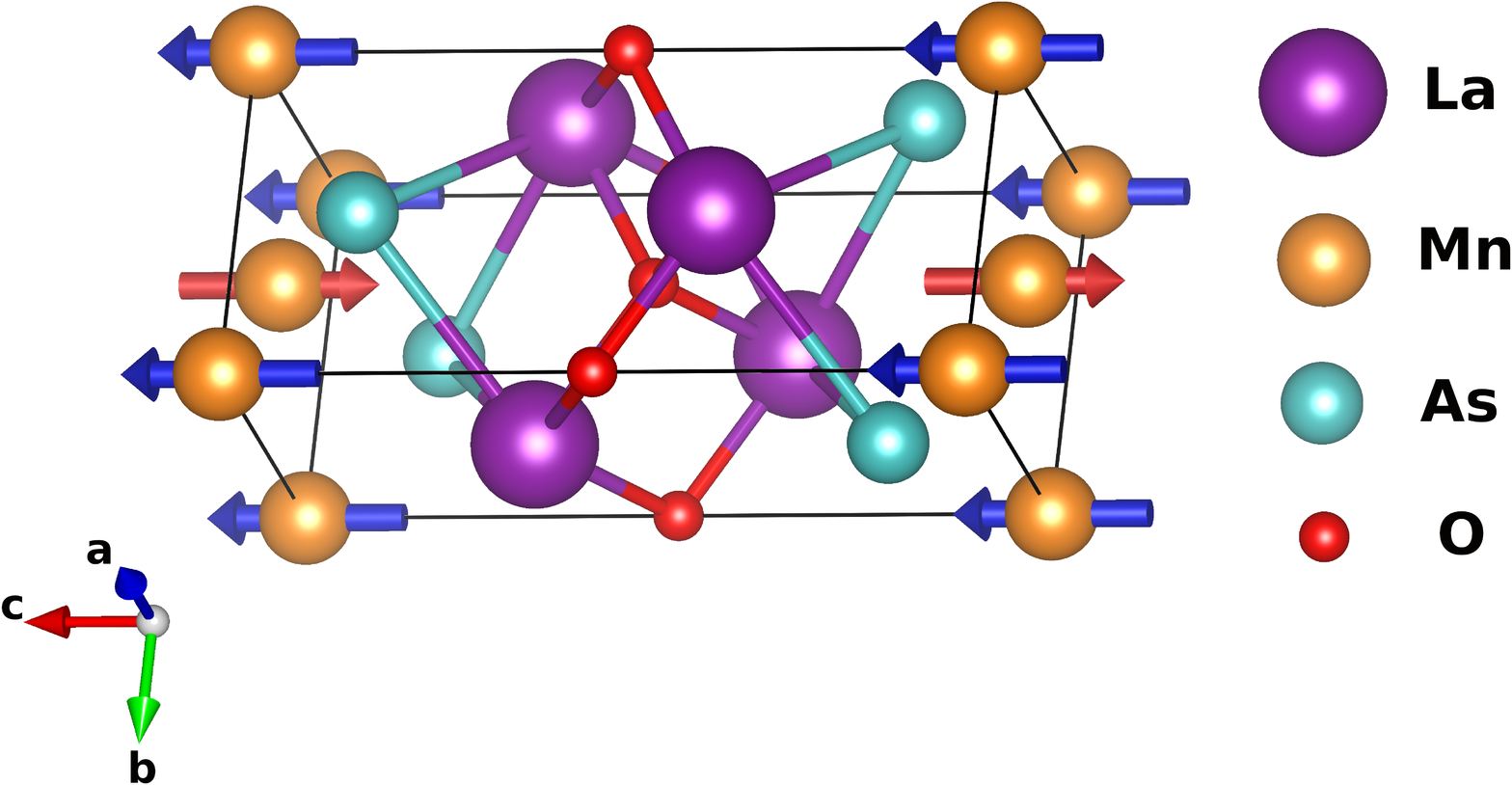}
          \caption{LaMnAsO}
          \label{fig:Structure_LaMnAsO}
     \end{subfigure}
     \centering
    \begin{subfigure}[b]{0.39\textwidth}
         \centering
         \includegraphics[width=\textwidth]{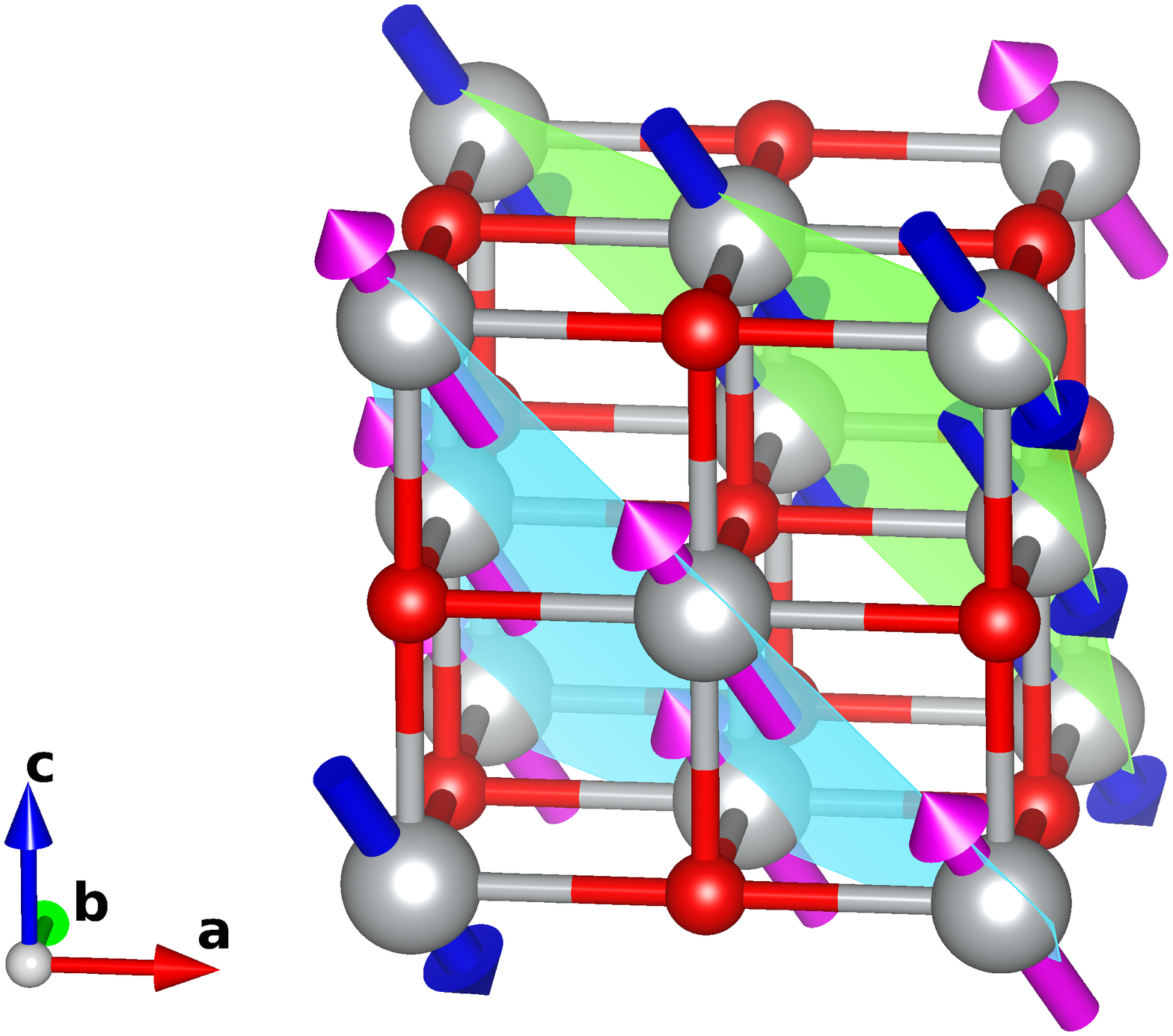}
         \caption{NiO}
        \label{fig:Structure_NiO}
    \end{subfigure}
    \caption{Crystal structure of (a) LaMnAsO (space group $P$4/$nmm$) and (b) NiO (space group $Fm\bar{3}m$) with directions of the magnetic moments along one of the easy directions in the light-green and light-blue colored (111)-planes.}
\end{figure}

In this work, we use a multislice simulation framework based on the paraxial Pauli equation \citep{edstrom_elastic_2016, edstrom_magnetic_2016} to explore the dependence of the intensity of antiferromagnetic Bragg spots both in LaMnAsO (space group $P4/nmm$) and NiO (space group $Fm\bar{3}m$) on sample thickness and acceleration voltage. The lattice parameters of LaMnAsO were set to $a=b=4.12$~\AA{}, $c=9.05$~\AA{} and for NiO $a=b=c=4.17$~\AA{} \citep{mcguire_short-_2016, cairns_xray_1933}.
For a selected sample thickness of 123~nm and an acceleration voltage equal to 300~kV, these simulations were also performed with the explicit inclusion of atomic vibrations, leading to thermal diffuse scattering (TDS), in order to compare the intensity of magnetic Bragg spots to the intensity of the TDS background.

Our motivation in choosing NiO for our computational study is largely motivated by the experimental results obtained by Loudon \cite{loudon_antiferromagnetism_2012} and the questions we posed earlier. The choice of the second antiferromagnetic material LaMnAsO in our calculations was motivated by the presence of heavier elements in this compound (La and As), and as a result of that, we expect stronger TDS, potentially making detection of a weak Bragg spot more challenging. In this way, we can qualitatively benchmark the feasibility to detect the magnetic Bragg spot in experiments across a wide range of material compositions and experimental conditions.

%
%
%
%
\begin{figure*}
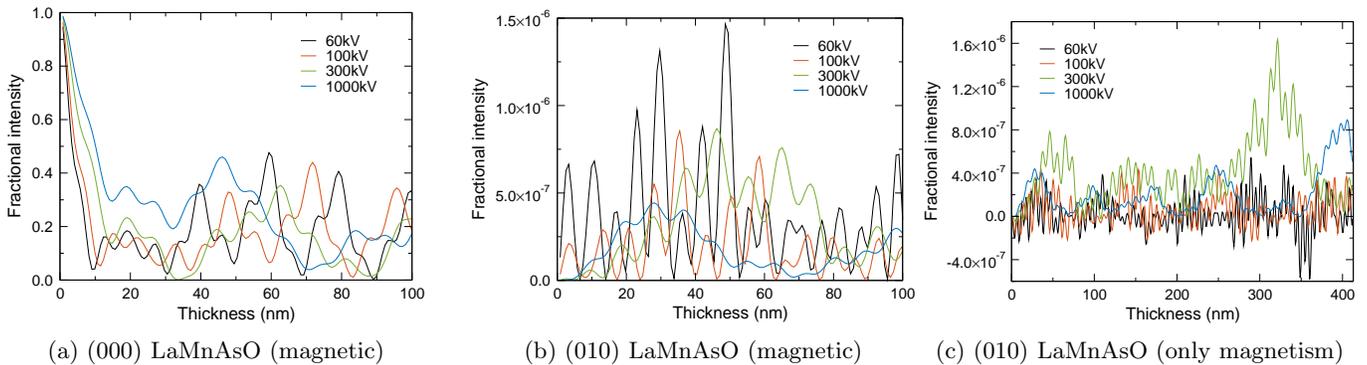

     \centering
     \begin{subfigure}[b]{0.31\textwidth}
         \centering
         \includegraphics[clip, width=\textwidth]{LaMnAsO_000_Magnetic.eps}
          \caption{(000) LaMnAsO (magnetic)}
          \label{fig:Volt000LaMnAsO}
     \end{subfigure}
     \hfill
     \begin{subfigure}[b]{0.33\textwidth}
         \centering
         \includegraphics[clip, width=\textwidth]{LaMnAsO_110_Magnetic.eps}
         \caption{(010) LaMnAsO (magnetic)} 
         \label{fig:Volt110LaMnAsO}
    \end{subfigure}
         \centering
     \begin{subfigure}[b]{0.32\textwidth}
         \centering
         \includegraphics[clip, width=\textwidth]{LaMnAsO_110_DIFF.eps}
         \caption{(010) LaMnAsO (only magnetism)}
         \label{fig:LaMnAsO_DIFF}
    \end{subfigure}
    \caption{Voltage and thickness dependence in the static model of (a) the intensity of the direct beam (000), and (b) the magnetic Bragg spots (010) in LaMnAsO. (c) The same as (b), but with subtracted forbidden beam reflections from non-magnetic calculations.}
\end{figure*}
\begin{figure}
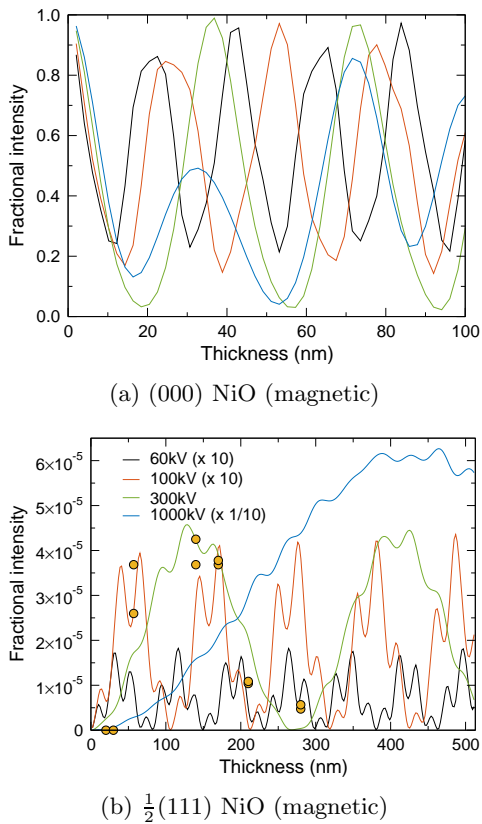

\centering
     \begin{subfigure}[b]{0.35\textwidth}
         \centering
         \includegraphics[clip, width=\textwidth]{NiO_000_Magnetic.eps}
          \caption{(000) NiO (magnetic)} \phantom{hello world}
          \label{fig:Volt000NiO}
     \end{subfigure}
     \centering
     \begin{subfigure}[b]{0.35\textwidth}
         \includegraphics[clip, width=\textwidth]{NiO_Voltage_1_2_111_Magnetic2.eps}
         \caption{$\frac{1}{2}$(111) NiO (magnetic)} \phantom{hello world}
         \label{fig:Volt1_2_111NiO}
    \end{subfigure}
    \caption{Voltage and thickness dependence in the static model of (a) the intensity of the direct beam (000), and (b) the intensity of magnetic Bragg spots $\frac{1}{2}(111)$ in NiO. Scaled experimental data (circles) are reproduced from Ref.~\cite{loudon_antiferromagnetism_2012}.}
\end{figure}
\section{\label{sec:Computational details}COMPUTATIONAL DETAILS}


The Pauli multislice method used in this work was described in detail by Edström et al. \cite{edstrom_elastic_2016,edstrom_magnetic_2016} and is based on a paraxial approximation to Pauli's equation,
\begin{equation} \label{eq:pauli}
\begin{aligned}[b]
 \frac{\partial \Psi}{\partial z} = \frac{im}{\hbar}(\hbar k + eA_z)^{-1}\Bigg\{\frac{\hbar^2 \nabla_{xy}^2}{2m} + \frac{ie \hbar}{m}\mathbf{A}_{xy} \cdot \nabla_{xy}   \\
  - \frac{\hbar keA_{z}}{m} - \frac{e \hbar}{2m}\bm{\sigma} \cdot \mathbf{B} + eV\Bigg\}\Psi,
 \end{aligned}
\end{equation}
instead of Schrödinger’s equation used in the standard multislice method \cite{cowley_scattering_1957}. In this equation, $V$ is the electrostatic potential, $\nabla_{xy}$ is the in-plane gradient operator, $\Psi$ is a wave function including both spin up and down components (Pauli spinor), and $k$ and $m$ are the relativistic values of the wavenumber and mass of the electron, respectively \cite{kirkland_advanced_2010}. The wavefunction is normalized to one, therefore intensities shown in the plots below refer to a fraction of incoming beam intensity.

We note that magnetic Bragg spot intensities calculated for spin-up vs spin-down polarized electron beams differ negligibly. This is because typically the strongest contribution due to magnetism comes from the term $\frac{\hbar k e A_z}{m}$ \cite{loudon_antiferromagnetism_2012, lubk_differential_2015}, which is insensitive to the spin of the electron beam. Nevertheless, in our simulations we evaluate all terms according to Eq.~(\ref{eq:pauli}).

%
\begin{figure*}
     \begin{subfigure}[b]{0.32\textwidth}
         \centering
         \includegraphics[clip,width=\textwidth]{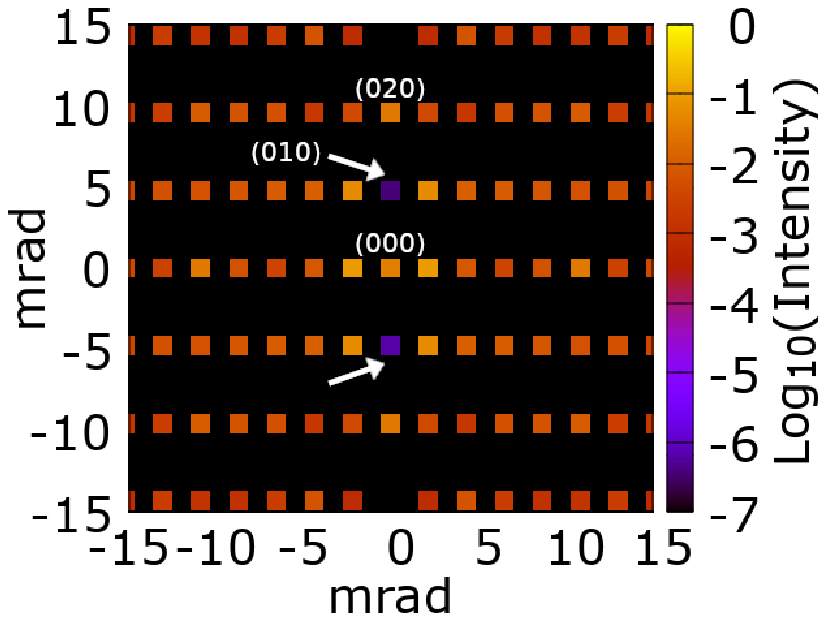}
         \caption{LaMnAsO (magnetic) }
         \label{fig:LaMnAsO_Magnetic}
    \end{subfigure}
         \hfill
     \begin{subfigure}[b]{0.334\textwidth}
         \centering
         \includegraphics[clip, width=\textwidth]{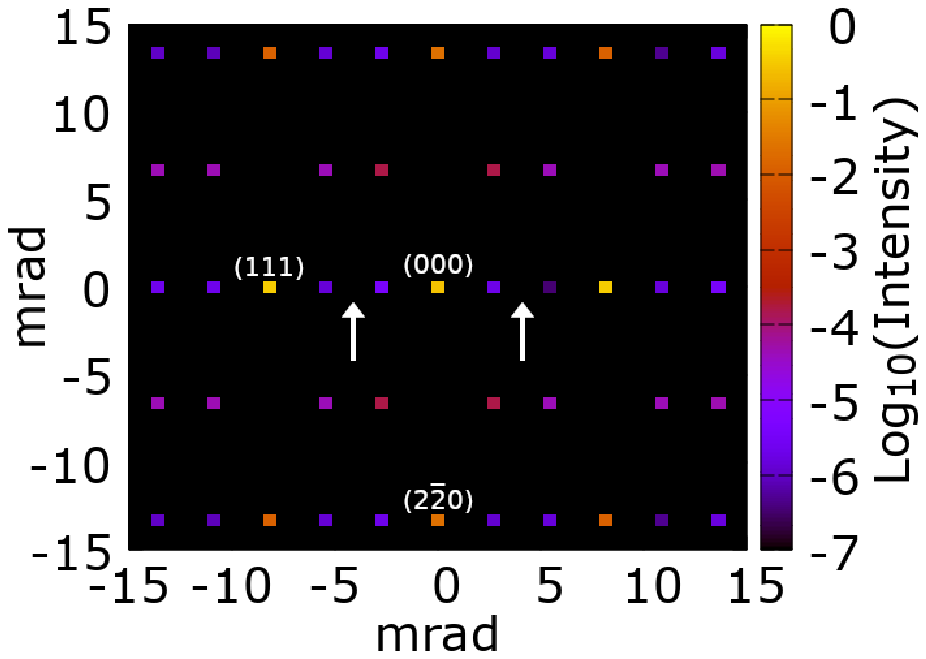}
         \caption{NiO (non-magnetic)}
         \label{fig:NiO_Non_magnetic}
    \end{subfigure}
             \hfill
     \begin{subfigure}[b]{0.32\textwidth}
         \centering
         \includegraphics[clip, width=\textwidth]{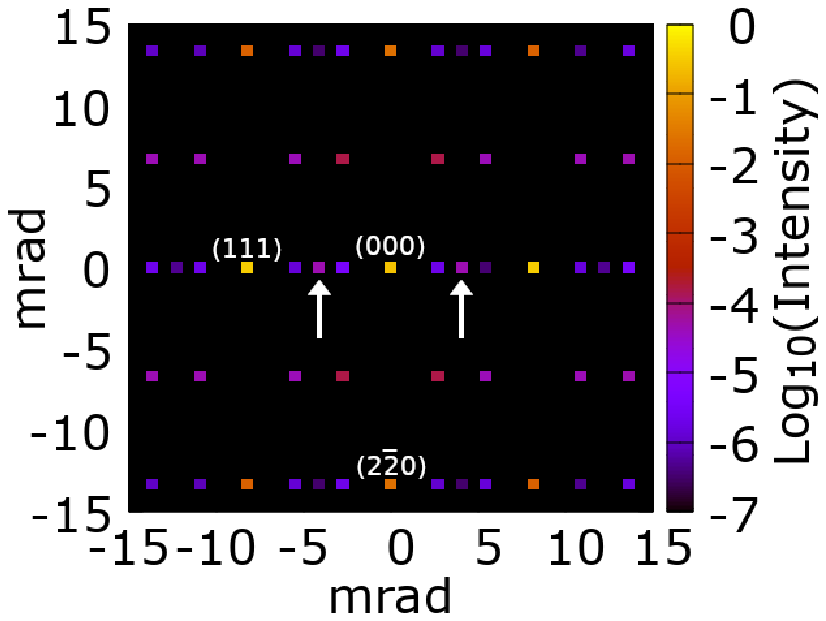}
         \caption{NiO (magnetic)}
         \label{fig:NiO_Magnetic}
    \end{subfigure}
    \caption{(a) Diffraction pattern for magnetic LaMnAsO at 300~kV and 123~nm thickness. Diffraction patterns at 300~kV and 123~nm thickness for (b) non-magnetic and (c) magnetic NiO.}
\end{figure*}
The microscopic magnetic vector potential $\mathbf{A(r)}$ and microscopic magnetic field $\mathbf{B(r)}$ were generated using a parametrization described by Lyon and Rusz \cite{lyon_parameterization_2021}. This process is based on the superposition of the microscopic magnetic fields and vector potentials, obtained by fitting a quasi-dipole model to the fields obtained via density functional theory for single atoms. It is worth underlining that for TDS calculations we are using Kirkland's parametrization \cite{kirkland_advanced_2010}, while for static model with included Debye–Waller factors (DWF) Peng's parametrization \cite{peng_robust_1996-1} is implemented. Kirkland's parametrization is using a combination of Gaussians and Lorentzians fitted against tabulated electron scattering data, while Peng's parametrization is based purely on Gaussians, which is particularly convenient for a real-space implementation of the DWF. The main difference between the parametrizations lies in an asymptotic behavior of electron form factors for large scattering angles \cite{lobato_accurate_2014} and that is not affecting our calculations, since the object of our interest \textemdash{} magnetic Bragg spots \textemdash{} are located at low scattering angles (below 10-20 mrad). The cut-off distance of atomic magnetic fields as well as atomic Coulomb potentials both in Kirkland's and Peng's parametrization was set to 4~\AA{}.

We have performed calculations both for a static and a vibrating lattice using a parallel beam (plane-wave) incident orthogonal to the direction of the magnetic moments. This is qualitatively motivated by the Lorentz force, which is maximal when the magnetic field is perpendicular to the beam current, see also Eq.~(\ref{eq:pauli}) in Ref.~\cite{loudon_antiferromagnetism_2012}. It should be noted that the computational method based on Eq.~(\ref{eq:pauli}) is generally applicable to describe elastic scattering of paraxial electrons in electric and magnetic fields, and also allows one to simulate other electron microscopy imaging techniques, such as the differential phase contrast microscopy \cite{krizek_atomically_2022, kohno_real-space_2022} or simulate propagation of vortex beams through magnetic materials \cite{edstrom_elastic_2016, edstrom_magnetic_2016}. 

In our calculations, we use the LaMnAsO and NiO structures shown in Figs.~\ref{fig:Structure_LaMnAsO} and \ref{fig:Structure_NiO}. The structures were oriented so that the incident beam propagating along the $z$-axis was orthogonal to the magnetic moment directions parallel to the $x$-axis in LaMnAsO and $y$-axis in NiO, respectively. For LaMnAsO, this corresponds to a beam direction $[100]$, while for NiO it is $[11\bar{2}]$. The LaMnAsO compound, with space group $P$4/$nmm$ and N\'eel temperature of $T_{N}=360$~K, contains magnetic moments only on Mn atoms parallel to $c$-axis in a so called C-type antiferromagnetic structure \citep{mcguire_short-_2016} (magnetic space group $P4'/n'mm'$ \footnote{McGuire and Garlea \cite{mcguire_short-_2016} report a different magnetic space group $P4'/n'm'm$, but this is likely a typo, because symmetry operations of $P4'/n'm'm$ are not consistent with C-type antiferromagnetic LaMnAsO having Mn magnetic moments parallel to $c$-axis}), while the NiO (magnetic space group $C_{c}$2/$c$%
) contains an antiferromagnetic distribution of the spin magnetic moments only on Ni atoms. We consider 2.4~$\mu_B$ \citep{emery_variable_2011} for the magnetic moment of Mn atoms, and 1.7~$\mu_B$ \citep{kwon_unquenched_2000} for Ni atoms. NiO has a relatively high N\'eel temperature of $T_{N}=523$~K \cite{roth_magnetic_1958}. Directions of magnetic moments in NiO were assumed parallel and anti-parallel to $[1\bar{1}0]$ direction, which corresponds to a spin-flopped configuration \cite{loudon_antiferromagnetism_2012}. 

Static lattice calculations, where the crystal potential is smeared by DWF, were done both with and without magnetic fields (the latter implemented by setting the $\mathbf{A}$- and $\mathbf{B}$-fields to zero). In the text below, we refer to these calculations as ``static model''. 
We have done calculations for acceleration voltages of 60, 100, 300, and 1000~kV, respectively. These calculations serve as a qualitative pre-screening of the intensity of the magnetic Bragg spot as a function of acceleration voltage and sample thickness. 

LaMnAsO static model calculations were carried out on a grid $184 \times 84 \times 16$ per $9.05 \times 4.12 \times 4.12$~\AA$^3$ unit cell repeated $6 \times 13 \times 300$ in the $x, y$, and $z$ directions. For NiO the grid was set to $288 \times 120 \times 48$ pixels per orthogonal supercell of dimensions $14.46 \times 5.90 \times 10.23$~\AA$^3$, which was repeated $2 \times 5 \times 120$ times in the corresponding directions. 


To simulate the thermal diffuse scattering (TDS), we have done calculations explicitly including atomic displacements as well as magnetic interactions. These calculations treat both phenomena (vibrations and magnetism) simultaneously and allow for a quantitative comparison of intensities of magnetic Bragg spots and the TDS background, thus offering a quantitative way to assess the feasibility of detecting the magnetic Bragg spots in terms of signal-to-noise ratios (SNR). We assume that the magnetic moments in these calculations remain ordered and oriented perpendicular to the beam, i.e., the effect of temperature on the magnetic moments is not included.

\begin{table}
\caption{\label{tab:Displacements} Experimentally measured mean squared displacement (MSD) at room temperature for LaMnAsO by x-ray powder diffraction \cite{emery_variable_2011} and for NiO by neutron powder diffraction \cite{lee_magnetoelastic_2016}.}
\begin{tabular}{lrrrrr}
 \hline
 \hline	
 \multicolumn{2}{c}{LaMnAsO}        &        \multicolumn{2}{c}{NiO}\\
\hline	

Atom        &   MSD [\AA$^2$]  	            &   Atom	&   MSD [\AA$^2$]    \\
\hline				
La	        &	0.589		    &	Ni	&	0.367  \\
Mn	        &	0.740	        &	O	&	0.494	 \\
As	        &	0.680	        &		&		          \\
O	        &	0.494	        &		&                  \\
\hline
\hline
\end{tabular}
\end{table}  


To simulate diffraction patterns, we use an approach based on quantum excitation of phonon (QEP) theory \citep{forbes_quantum_2010} within the Einstein model, where an inelastic signal due to atomic vibrations can be obtained by sampling over possible atomic displacement configurations. The mean squared displacements for each atom in the two compounds analyzed here are specified in Table~\ref{tab:Displacements} according to experimental data. 

We have generated 250 snapshots with normally distributed (Gaussian) random displacements according to experimental values of MSD. Averaging over intensities of exit wave functions calculated for each of these snapshots results in an incoherent scattering intensity. The exit wave functions were obtained by the paraxial Pauli equation, Eq.~(\ref{eq:pauli}) above.

The supercell for TDS calculations for LaMnAsO was the size of  $18.10 \times 16.49 \times 1236.75$~\AA$^3$ containing 19200 atoms, and for NiO $28.93 \times 29.52 \times 1227.19$~\AA$^3$ containing 115200 atoms. Dimensions of these supercells differ from their static counterparts in order to avoid artificial periodicity along the $z$-axis, while keeping the computing costs manageable.

On the same structure models, we have also done calculations excluding magnetic fields, but with atomic vibrations included, in order to estimate the net TDS background.


\section{\label{sec:Results}Results}

\begin{figure*}
     \centering
     \begin{subfigure}[b]{0.32\textwidth}
         \centering
         \includegraphics[width=\textwidth]{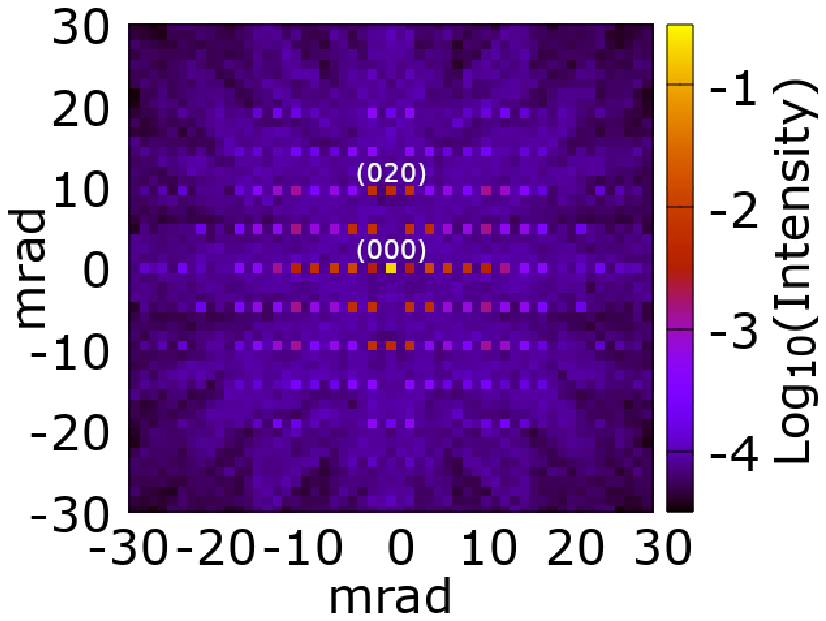}
         \caption{LaMnAsO Non-Magnetic}
         \label{fig:LaMnAsO_TDS_LOG_2}
    \end{subfigure}
         \hfill
     \begin{subfigure}[b]{0.32\textwidth}
         \centering
         \includegraphics[width=\textwidth]{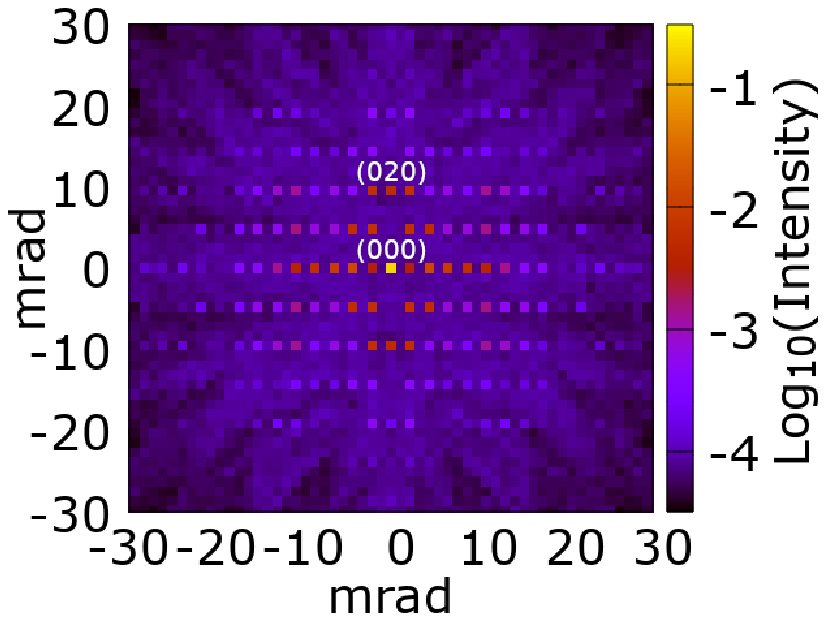}
         \caption{LaMnAsO Magnetic}
         \label{fig:LaMnAsO_TDS_Magnetic}
    \end{subfigure}
             \hfill
     \begin{subfigure}[b]{0.32\textwidth}
         \centering
         \includegraphics[width=\textwidth]{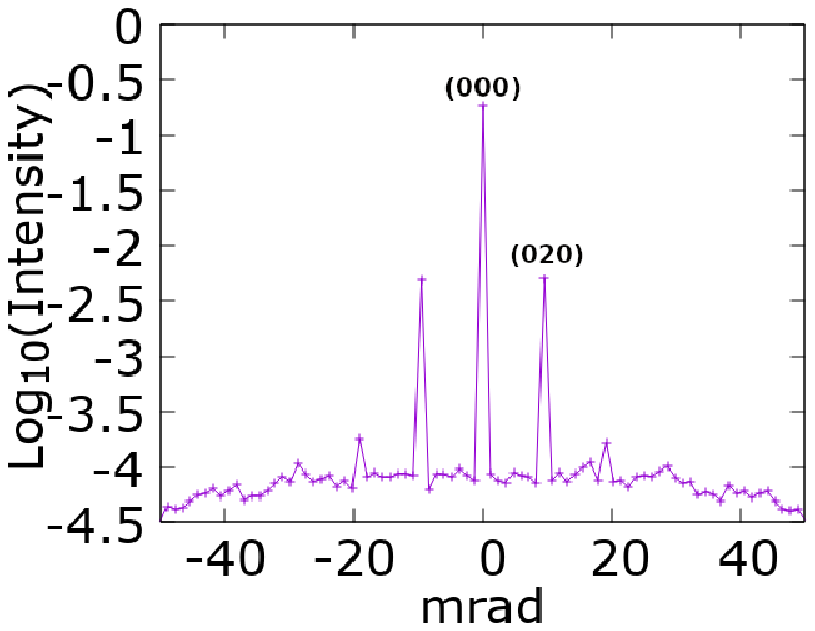}
         \caption{LaMnAsO Magnetic}
         \label{fig:LaMnAsO_TDS_Magnetic_LINE}
    \end{subfigure}
         \centering
     \begin{subfigure}[b]{0.32\textwidth}
         \centering
         \includegraphics[width=\textwidth]{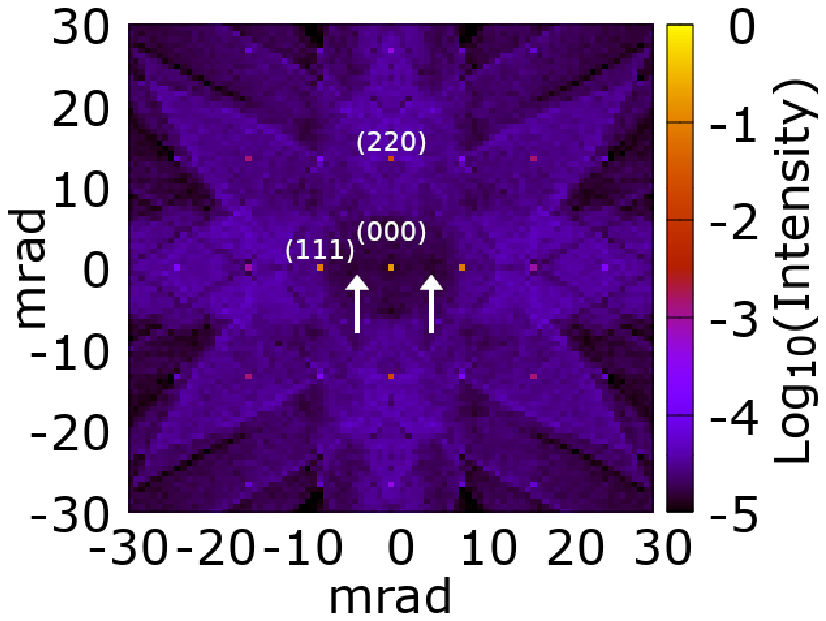}
         \caption{NiO Non-Magnetic}
         \label{fig:NiO_TDS_LOG_2}
    \end{subfigure}
         \hfill
     \begin{subfigure}[b]{0.32\textwidth}
         \centering
         \includegraphics[width=\textwidth]{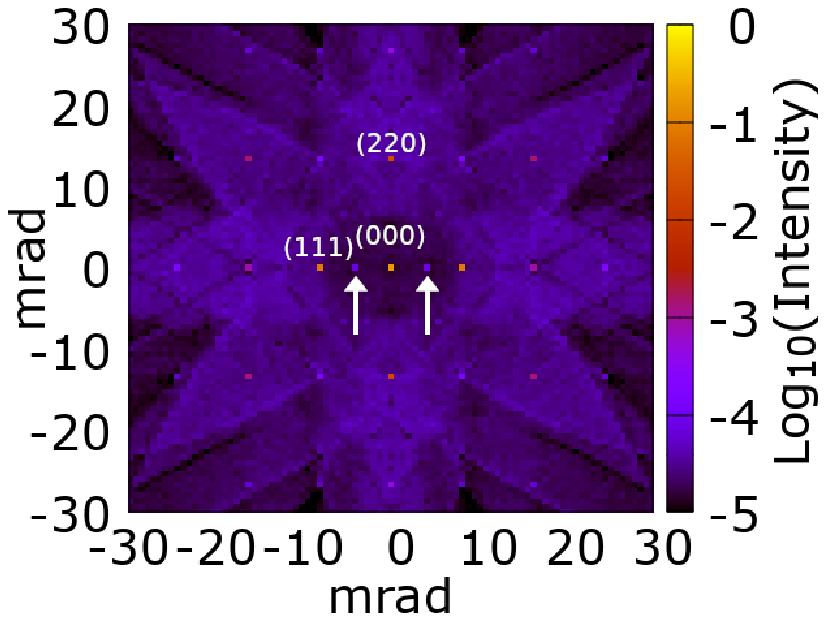}
         \caption{NiO Magnetic}
         \label{fig:NiO_TDS_Magnetic}
    \end{subfigure}
             \hfill
     \begin{subfigure}[b]{0.32\textwidth}
         \centering
         \includegraphics[width=\textwidth]{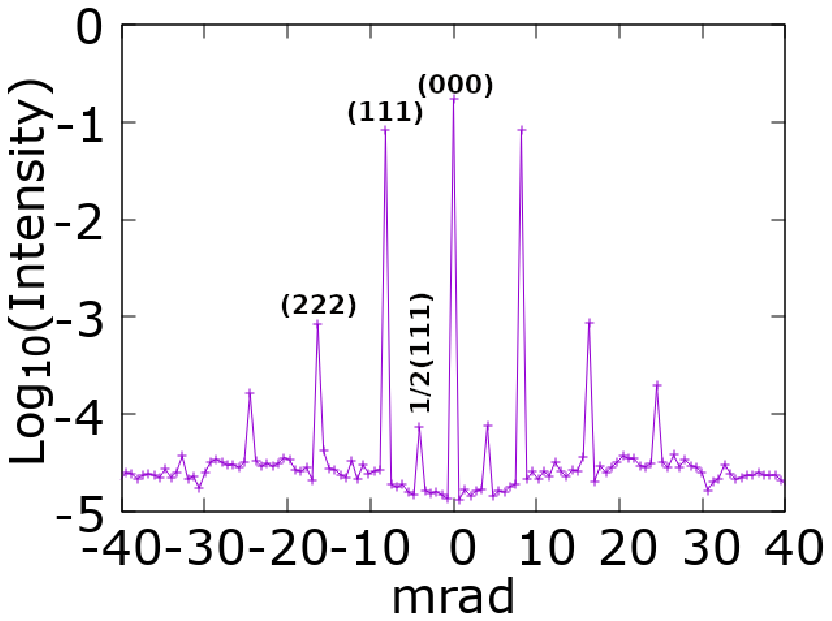}
         \caption{NiO Magnetic}
         \label{fig:NiO_TDS_Magnetic_LINE}
    \end{subfigure}
       
    \caption{QEP calculations with thermal diffuse scattering calculated for room temperature in (a-c) LaMnAsO and (d-f) NiO. The left column, panels (a) and (d), show diffraction patterns calculated with zero magnetic fields. Panels (b) and (e) show results with magnetic fields included. Plots (c) and (f) show linear profiles through the diffraction patterns calculated with magnetic fields taken at $\theta_x=0$~mrad for LaMnAsO and $\theta_y=0$~mrad for NiO, respectively.}
    \label{fig:NiO_LaMnAsO_TDS}
\end{figure*}
\subsection{\label{sec:Voltage and thickness}Static model calculations of thickness and voltage dependence of the intensity of magnetic Bragg scattering}
In the first step of our calculations, we have considered the static model including magnetism and tabulated Coulomb potentials smeared by DWF. 
As mentioned in the previous section, we have tested the influence of the acceleration voltage in the range from 60 to 1000~kV on the intensity of magnetic Bragg scattering and its thickness dependence. 

The calculated intensities of the direct beam and at the position of antiferromagnetic Bragg spots are summarized in Figs.~\ref{fig:Volt000LaMnAsO}, \ref{fig:Volt110LaMnAsO}, \ref{fig:Volt000NiO}, and \ref{fig:Volt1_2_111NiO}. For clarity, these results are shown only for samples up to 100~nm thick, except for panel Fig.~\ref{fig:Volt1_2_111NiO}, where we include comparison with experiments. Particularly in Fig.~\ref{fig:Volt110LaMnAsO} we see intense oscillations with a relatively short period in thickness. They are remarkably strong at the lowest acceleration voltage of 60~kV. However, these intensities can not be solely ascribed to a magnetic signal. One should note that in the static model calculations, at the position of magnetic Bragg spots there will generally be a nonzero intensity contribution due to forbidden reflections. Structurally forbidden reflections can have a nonzero intensity in diffraction pattern, when several atomic planes along the beam direction should contribute to achieve the destructive interference \cite{nishio_multi-slice_1994}. The reason lies in slight changes of the electron beam wave function when propagating through each of these atomic planes. When studying such weak signals as the antiferromagnetic Bragg spots, contributions of forbidden reflections can be, in relative terms, non-negligible. Therefore, we have also recalculated the static model without considering magnetism, and subtracted the obtained forbidden beam reflection intensities from those obtained by magnetic calculations. The results for LaMnAsO (Fig.~\ref{fig:LaMnAsO_DIFF}; note the extended thickness range) now show a dramatically different picture. The difference signal becomes strongest for higher acceleration voltages. Occasional negative intensities show that considering the magnetic Bragg peak intensity as an additive signal can only be done approximately. For NiO, the forbidden beam reflections at the position of the magnetic Bragg spot were of much lower intensity than the magnetic signal, so the results in Fig.~\ref{fig:Volt1_2_111NiO} remained visually unchanged by taking the difference (figure not shown).

As can be seen in Fig.~\ref{fig:Volt1_2_111NiO}, the magnetic Bragg spot intensity in NiO is significantly lower at acceleration voltages of 60~kV and 100~kV, when compared to 300~kV and, particularly, 1000~kV. However, at 1000~kV, the intensity of the magnetic Bragg spot becomes high only for very large sample thicknesses, which is a disadvantage from the practical point of view. Overall, our results support the choice made by Loudon to use 300~kV acceleration voltage for NiO. Moreover, we note that our simulations at 300~kV show oscillations with the main period of approximately 265~nm, which is in good agreement with experimental data of Loudon \cite{loudon_antiferromagnetism_2012}.

For LaMnAsO the difference between the magnetic and non-magnetic intensity of the Bragg spot, Fig.~\ref{fig:LaMnAsO_DIFF}, suggests that the acceleration voltage of 300~kV provides an advantage also here. Comparing with Fig.~\ref{fig:Volt110LaMnAsO}, we note that the intensity  of forbidden reflections decreases as the acceleration voltage increases, which can be qualitatively understood, because the modification of the electron beam wave function by a single atomic plane reduces with increasing acceleration voltage. 

Figures~\ref{fig:Volt000LaMnAsO} and \ref{fig:Volt000NiO} show, for reference, the thickness dependence of the direct beam $(000)$ in both compounds. Overall, we can see that in relative terms, the magnetic Bragg spot is less intense in LaMnAsO in comparison to NiO. This is likely due to the presence of heavier elements in LaMnAsO, which leads to a stronger Coulomb potential in comparison to NiO and therefore weaker magnetic fields in the relative sense.

Figures~\ref{fig:NiO_Magnetic} and \ref{fig:LaMnAsO_Magnetic} show diffraction patterns of the two compounds, including relevant Miller indices and expected positions of magnetic Bragg spots. For NiO we observe in Fig.~\ref{fig:NiO_Magnetic} a number of additional spots due to forbidden beam reflections. However, as Fig.~\ref{fig:NiO_Non_magnetic} shows using the calculation excluding magnetic fields, the forbidden beam reflections at the position of the magnetic Bragg spots are of much lower intensity (as was pointed out above) and thus not visible at the chosen color scale.

From the static model calculations, we conclude that in order to detect the magnetic Bragg spots, it appears to be advantageous to use higher acceleration voltages and sample thicknesses of around 100~nm. Motivated by these results, we have chosen 300~kV acceleration voltage and a sample thickness of 123~nm for both systems in calculations that include the atomic displacements.

\begin{figure}
     \centering
     \begin{subfigure}[b]{0.40\textwidth}
         \centering
         \includegraphics[width=\textwidth]{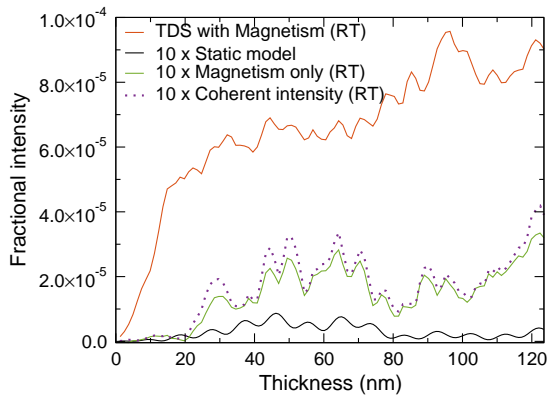}
         \caption{LaMnAsO Magnetic}
         \label{fig:LaMnAsO_TDS_RT_DEP}
    \end{subfigure}
      \hfill
     \begin{subfigure}[b]{0.4\textwidth}
         \centering
         \includegraphics[width=\textwidth]{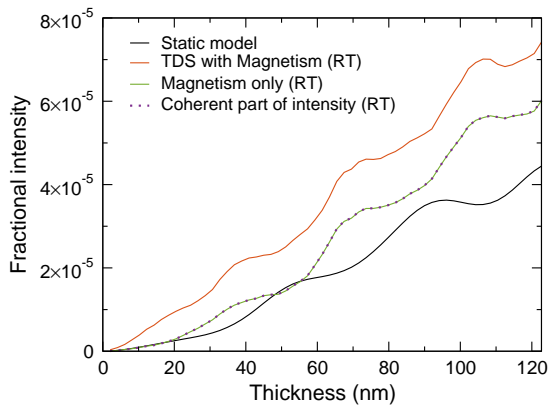}
         \caption{NiO Magnetic}
         \label{fig:NiO_TDS_DEP}
    \end{subfigure}
        
    \caption{ Plots (a) and (b) show the intensity of the antiferromagnetic Bragg spot for LaMnAsO and NiO, respectively, as a function of the thickness in the static model (black line), from calculations with atomic displacements (orange line), and from the magnetic contribution in the TDS calculations (green line), where the TDS intensity was subtracted. The dark dotted line is obtained as the coherent intensity at the magnetic Bragg spot, calculated from the averaged exit function.}
    \label{fig:NiO_LaMnAsO_TDS_2}
\end{figure}

\subsection{\label{sec:Frozen Phonon} Non-magnetic TDS simulations of NiO and LaMnAsO}
As a reference for calculations that consider magnetism and atomic vibrations simultaneously, we have performed TDS calculations within Einstein’s model approximation, including tabulated Coulomb potentials with zero magnetic fields for both compounds. 


TDS calculations are important in the context of estimating the influence of atomic motion and its resulting effects on the diffraction pattern, including the intensities of Bragg spots. Taking this fact into account will help in defining the actual intensity of the antiferromagnetic Bragg spot in relation to the thermal effects overlapping with this weak phenomenon. Knowing these values will also be needed to calculate the acquisition time under which the sample will need to be exposed to the incident electron beam, such that the transmitted electron counts are sufficient to distinguish the antiferromagnetic Bragg spot from the TDS background in the diffraction pattern.

In the case of TDS calculations (at 123~nm and 300~kV) for the LaMnAsO compound, shown in Fig.~\ref{fig:LaMnAsO_TDS_LOG_2}, the magnetic $(010)$ reflection is not visible (as expected in these non-magnetic calculations).
Similarly, the TDS calculations for NiO are shown in Fig.~\ref{fig:NiO_TDS_LOG_2}, where we can see that without magnetism the $\frac{1}{2}(111)$ reflection is not visible. Both calculations show well-resolved Kikuchi patterns arising from TDS at a relatively large sample thickness. The intensities of the direct beam $(000)$ in these calculations are 0.13 and 0.18 for LaMnAsO and NiO, respectively, with the total intensity normalized to 1.

%
\begin{figure*}
     \centering
     \begin{subfigure}[b]{0.32\textwidth}
         \centering
         \includegraphics[width=\textwidth]{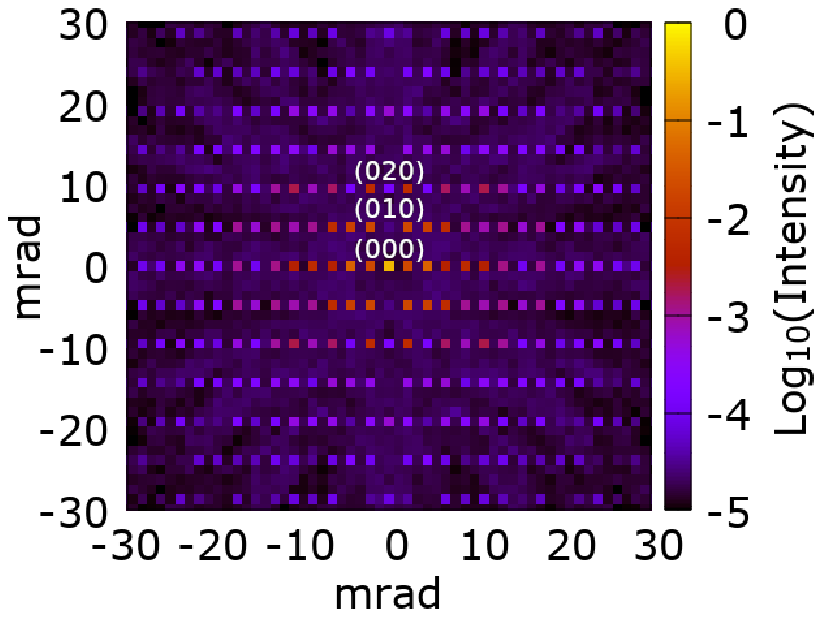}
          \caption{LaMnAsO (LT)}
          \label{fig:LaMnAsO_TDS_1_10_MSD}
     \end{subfigure}
     \hfill
     \begin{subfigure}[b]{0.32\textwidth}
         \centering
         \includegraphics[width=\textwidth]{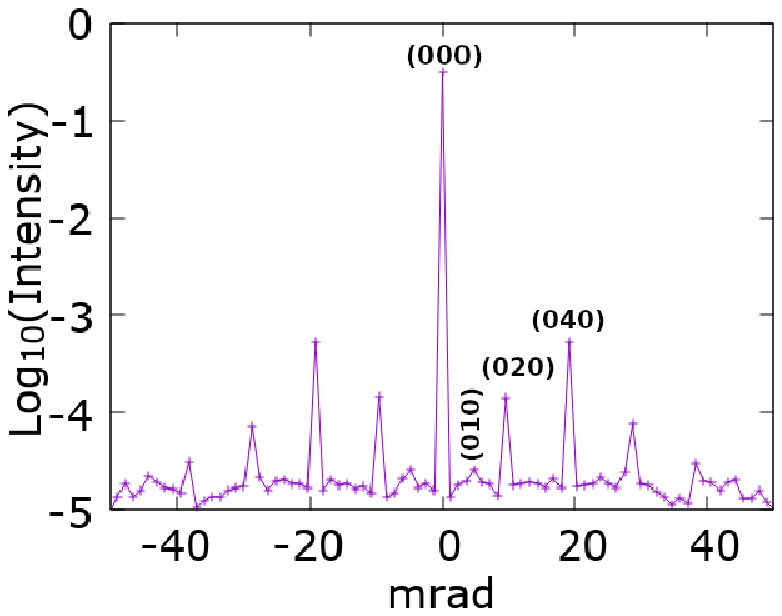}
         \caption{LaMnAsO (LT)}
         \label{fig:LaMnAsO_TDS_LINE_1_10_MSD}
    \end{subfigure}
         \hfill
    \begin{subfigure}[b]{0.31\textwidth}
         \centering
         \includegraphics[width=\textwidth]{LaMnAsO_LT_DWF_PSIK.eps}
          \caption{LaMnAsO (LT)}
          \label{fig:LaMnAsO_TDS_1_10_MSD_DEP}
     \end{subfigure}
    \caption{Panels (a-c) replicate Fig.~\ref{fig:NiO_LaMnAsO_TDS} panels (b-c) and Fig.~\ref{fig:LaMnAsO_TDS_RT_DEP}, calculated for 10-times reduced mean squared displacement of all atoms, a proxy for low-temperature (LT) calculations.}
    \label{fig:LaMnAsO_TDS_Low_Temp}
\end{figure*}
\subsection{\label{sec:Frozen Phonon_Magnetic} TDS simulations of NiO and LaMnAsO including magnetic fields}
In this section, we present TDS calculations within Einstein’s model approximation, including both electrostatic potentials and magnetic fields.

In the case of NiO we can observe the antiferromagnetic Bragg spot appearing in the $\frac{1}{2}(111)$ spot, see Fig.~\ref{fig:NiO_TDS_Magnetic}. The total intensity of the antiferromagnetic Bragg spot in NiO (at 123~nm and 300~kV) is  7.42$\times$10$^{-5}$, meaning that approximately 7 electrons for every 10000 are scattered in the $\frac{1}{2}(111)$ direction. The above value includes the TDS background, whose value was estimated to be 1.41$\times$10$^{-5}$ from a non-magnetic calculation, and 1.51$\times$10$^{-5}$, within the magnetic calculation from neighboring pixels in the diffraction pattern. These two values are close to each other, as could be expected. Consequently, the  signal-to-noise ratio (SNR) for beam current 1~pA and time of the sample exposure 0.1~s is estimated to be above 12, this high value suggesting that the detection of magnetic Bragg spot in NiO at room temperature and a suitable thickness and acceleration voltage is feasible.

In Fig.~\ref{fig:NiO_TDS_DEP} we show the comparison of the thickness dependence in TDS and static model calculations. The two curves show a semi-quantitative agreement, supporting the use of static calculations for a pre-screening of the dependence of the intensity of magnetic Bragg spot on acceleration voltage and sample thickness. In addition, we have used the exit wavefunctions calculated for all structure snapshots to construct an averaged wavefunction. According to the QEP model \cite{forbes_quantum_2010}, the averaged exit wavefunction represents a coherent elastic scattering and thus its amplitude squared offers an alternative route to extract the intensity of the magnetic Bragg spot. The result is shown using a dark dotted line and for NiO it is in a very close agreement with the intensity extracted as a difference of magnetic and non-magnetic TDS calculation. The agreement remains excellent also in the LaMnAsO case discussed below. 

In the LaMnAsO case, the (010) antiferromagnetic Bragg spot at room temperature is thoroughly covered with TDS intensity, which is stronger than in NiO due to the presence of heavier elements such as La and As in its composition, see Fig.~\ref{fig:LaMnAsO_TDS_Magnetic}. Neither the diffraction pattern nor the linear profile passing through the (010) spots (Fig.~\ref{fig:LaMnAsO_TDS_Magnetic_LINE}) reveals any peak at the (010) position.

For this reason, we have performed calculations with artificially reduced TDS by decreasing the MSD values by a factor of 10 for each atom as a simplified model of LaMnAsO at a low temperature. In the harmonic approximation and classical statistical mechanics, that would correspond to a temperature of 30~K. However, this estimate neglects the influence of nuclear quantum effects \cite{lofgren_influence_2016}, which eventually become the main contribution to MSD at the lowest temperatures. Based on the actual strength of nuclear quantum effects in LaMnAsO (evaluation of which is outside the scope of this work), the ten-fold reduced MSD would correspond to a temperature below 30~K, though potentially such low MSD might even not be reachable at absolute zero. For the purposes of this work, we consider the ten-fold reduced MSDs as a model for exploring effects of reduced temperature. Results are summarized in Fig.~\ref{fig:LaMnAsO_TDS_Low_Temp}.

The resulting reduction of TDS led to an observation of the (010) antiferromagnetic Bragg spot with the total intensity (including the TDS background) equal to 2.64$\times$10$^{-5}$ for the sample thickness 123~nm, making it visible on top of the TDS background intensity. The TDS intensity of 2.12$\times$10$^{-5}$ was extracted from a non-magnetic calculation. The value is again similar to the average intensity of the surrounding pixels (1.93$\times$10$^{-5}$) in a magnetic calculation. Based on these values, at the same conditions as above (1~pA beam current and 0.1~s dwell time) the SNR would be only slightly above 1. Raising the beam current to 50~pA and keeping the sample exposure time at 0.1~s, the SNR would become larger than 8. Overall, the simulations suggest that the detection of a magnetic Bragg spot in LaMnAsO is substantially more challenging, requiring reduced temperatures, larger beam currents, and/or extended acquisition times than a similar experiment in NiO.

Returning to the room temperature calculation, we can make a simplifying assumption that the magnetic contribution to the intensity of the antiferromagnetic Bragg spot will be approximately the same as in the calculation with reduced MSD \textemdash{} in this way, we would obtain a magnetic signal intensity of $0.53 \times 10^{-5}$. For the TDS, we obtain from a non-magnetic calculation $8.7 \times 10^{-5}$ at this scattering angle. At 1~pA beam current and 0.1~s exposure time, this would lead to a very low SNR of 0.45, well below any detection criterion. One could make the signal detectable by, for example, increasing the beam current to 100~pA, by which the SNR would be raised to about 4.5.


In Fig.~\ref{fig:LaMnAsO_TDS_1_10_MSD_DEP} we compare the thickness dependence of the magnetic Bragg spot intensity obtained in calculations including TDS as well as those from static model calculations. As in Figs.~\ref{fig:LaMnAsO_TDS_RT_DEP} and \ref{fig:NiO_TDS_DEP}, we see that the static calculation semi-quantitatively agrees with the calculation including TDS. The agreement between static and QEP calculation is better here than in the aforementioned Figures, which is likely linked to a lower total TDS intensity, since these electrons are effectively removed from the elastic channel and the inclusion of such effect on the intensities of Bragg reflections would require the use of absorptive potentials in the static calculation \cite{weickenmeier_computation_1991}. On the other hand, the QEP calculation and the difference of magnetic and non-magnetic TDS intensities are both based on a set of calculations with atomic displacements, where the absorption effects are implicitly included. This likely explains their close agreement.



%
\section{\label{sec:Summary}SUMMARY AND CONCLUSIONS}
We have performed multislice simulations based on the paraxial Pauli equation to investigate the influence of magnetic properties, constituents of the material, and other experimental parameters including acceleration voltage and sample thickness, in antiferromagnetic materials (LaMnAsO and NiO) on their electron diffraction patterns, which can be observed experimentally using transmission electron microscopes. 

From our calculations, we observed that for NiO it is quite possible to observe the $\frac{1}{2}$(111) antiferromagnetic Bragg spot at room temperature as was shown experimentally by Loudon. We have verified that the antiferromagnetic Bragg spot intensity is significantly stronger than the thermal diffuse scattering intensity. Our calculations also provide a good agreement with the measured thickness dependence of the magnetic Bragg spot intensity.

Our simulations for LaMnAsO, containing heavier elements than NiO, suggested that for systems with strong thermal diffuse scattering it can be challenging to detect the magnetic Bragg spots. For such systems, it might be necessary to work at reduced temperatures and/or to perform data acquisition for an extended time with a sufficiently high beam current. The results also indicate the need to select an appropriate acceleration voltage, which for the materials studied here is found to be 300 kV. The simulation methods presented will be valuable in finding favorable experimental settings, paving the way to use TEM for high resolution detection of complex magnetic orders. 

\begin{acknowledgments}
J.S.-A. gratefully acknowledges financial support from the National Science Center Poland under decision DEC-2019/35/O/ST5/02980 (PRELUDIUM-BIS 1) and National Agency for Academic Exchange Poland under decision PPN/STA/2021/1/00014/U/00001. J.R., P.Z., J.A.C.R., and K.L. acknowledge Swedish Research Council, Olle Engkvist's foundation, and Carl Tryggers foundation for financial support. A.E. acknowledges Swedish Research Council (2018-06807 and 2022-04720) and the G\"oran Gustafsson foundation for financial support. Part of the calculations was enabled by resources provided by the Swedish National Infrastructure for Computing (SNIC), partially funded by the Swedish Research Council through grant agreement no. 2018-05973. M.W. acknowledges financial support from the National Science Center Poland under decision DEC-2018/30/E/ST3/00267 (SONATA-BIS 8).
\end{acknowledgments}


\bibliography{BraggScattering}

\end{document}